\documentclass[12pt]{article}
\usepackage{amsmath}
\usepackage{graphicx,psfrag,epsf}
\usepackage{enumerate}
\usepackage{natbib}
\usepackage{textcomp}
\usepackage[hyphens]{url} % not crucial - just used below for the URL
\usepackage{hyperref}

\newcommand{\blind}{0}

\newcommand{\nT}{\ensuremath{T}}
\newcommand{\ns}{\ensuremath{K}}
\newcommand{\no}{\ensuremath{N}}

\newcommand{\state}{\ensuremath{S}}
\newcommand{\fstate}{\ensuremath{s}}
\newcommand{\obs}{\ensuremath{Y}}

\newcommand{\modpar}{\ensuremath{\mbox{\boldmath$\theta$}}}

\newcommand{\dens}{\ensuremath{p}}

\newcommand{\Prob}{\ensuremath{p}}

\newcommand{\mat}[1]{\ensuremath{\mathbf{#1}}}
\newcommand{\gvc}[1]{\mbox{\boldmath$#1$}}

\usepackage{setspace}
\usepackage{amsmath}
\usepackage{amsfonts}
\usepackage{booktabs}
\usepackage{longtable}
\usepackage{array}
\usepackage{multirow}
\usepackage{wrapfig}
\usepackage{float}
\usepackage{colortbl}
\usepackage{pdflscape}
\usepackage{tabu}
\usepackage{threeparttable}
\usepackage{threeparttablex}
\usepackage[normalem]{ulem}
\usepackage{makecell}
\usepackage{xcolor}

\begin{document}

\def\spacingset#1{\renewcommand{\baselinestretch}%
{#1}\small\normalsize} \spacingset{1}

%%%%%%%%%%%%%%%%%%%%%%%%%%%%%%%%%%%%%%%%%%%%%%%%%%%%%%%%%%%%%%%%%%%%%%%%%%%%%%

\if0\blind
{
  \title{\bf Ignorable and non-ignorable missing data in hidden Markov
models}

  \author{
        Maarten Speekenbrink \thanks{The authors would like to thank Don
Hedeker for sharing the data of the Schizophrenia clinical trial.} \\
    Department of Experimental Psychology, University College London\\
     and \\     Ingmar Visser \\
    Department of Developmental Psychology, University of Amsterdam\\
      }
  \maketitle
} \fi

\if1\blind
{
  \bigskip
  \bigskip
  \bigskip
  \begin{center}
    {\LARGE\bf Ignorable and non-ignorable missing data in hidden Markov
models}
  \end{center}
  \medskip
} \fi

\bigskip
\begin{abstract}
We consider missing data in the context of hidden Markov models with a
focus on situations where data is missing not at random (MNAR) and
missingness depends on the identity of the hidden states. In
simulations, we show that including a submodel for state-dependent
missingness reduces bias when data is MNAR and state-dependent, whilst
not reducing accuracy when data is missing at random (MAR). When
missingness depends on time but not the hidden states, a model which
only allows for state-dependent missingness is biased, whilst a model
that allows for both state- and time-dependent missingness is not.
Overall, these results show that modelling missingness as
state-dependent, and including other relevant covariates, is a useful
strategy in applications of hidden Markov models to time-series with
missing data. We conclude with an application of the state- and
time-dependent MNAR hidden Markov model to a real dataset, involving
severity of schizophrenic symptoms in a clinical trial.
\end{abstract}

\noindent%
\vfill

\newpage
\spacingset{1.45} % DON'T change the spacing!

\hypertarget{introduction}{%
\section{Introduction}\label{introduction}}

There is relatively little work on dealing with missing data in hidden
Markov models. \citet{albert2000transitional},
\citet{deltour1999stochastic}, and \citet{yeh2010estimating} consider
missing data in observed Markov chains. \citet{paroli2002parameter}
consider calculation of the likelihood of a Gaussian hidden Markov model
when observations are missing at random. \citet{yeh2012intermittent}
discuss the impact of ignoring missingness when missing data is, and is
not, ignorable. They show that if missingness depends on the hidden
states, i.e.~missingness is state-dependent, this results in biased
parameter estimates when this missingness is ignored. However, they
offer no solution to this problem. The objective of this paper is to do
so. Our approach is related to the work of \citet{yu2003semimarkov} who
allowed for state-dependent missingness in a hidden semi-Markov model
with discrete (categorical) outcomes. Following \citet{bahl1983maximum},
their solution is to code missingness into a special ``null value'' of
the observed variable, effectively making the variable fully observed.
Here, we instead model missingness with an additional (fully observed)
indicator variable. This, we believe, is conceptually simpler, and makes
it straightforward to add additional covariates to model the probability
of missing values.

The remainder of this paper is organized as follows: We start with a
brief overview of hidden Markov models and the formal treatment of
ignorable and non-ignorable missing data as established by
\citet{rubin1976inference} and \citet{little2014statistical}, with a
focus on hidden Markov models. We then consider state-dependent
missingness in hidden Markov models, and show in simulation studies how
including a submodel for state-dependent missingness provides better
estimates of the model parameters. When data is in fact missing at
random, the model with state-dependent missingness is not fundamentally
biased, although care must be taken to include relevant covariates, such
as e.g.~time. We conclude with an application of the method to a real
dataset, involving severity of schizophrenic symptoms in a clinical
trial.

\hypertarget{hidden-markov-models}{%
\subsection{Hidden Markov models}\label{hidden-markov-models}}

Let \(\obs_{1:\nT} = (\obs_1,\ldots,\obs_\nT)\) denote a time series of
(possibly multivariate) observations, and let \(\modpar\) denote a
vector of parameters. A hidden Markov model associates observations with
a time series of hidden (or latent) discrete states
\(\state_{1:\nT} = (\state_1,\ldots,\state_\nT)\). It is assumed that
each state \(\state_t \in \{1, \ldots, \ns\}\) depends only on the
immediately preceding state \(\state_{1-t}\), and that, conditional upon
the hidden states, the observations \(\obs_t\) are independent:
\begin{align}
        \Prob(\state_{t}|\state_{1:t-1},\modpar) &= 
        \Prob(\state_{t}|\state_{t-1},\modpar), \quad t=2, 3, \ldots, \nT
        \label{eq:hmmstates}
        \\ \dens(\obs_{t}|\state_{1:t-1}, \obs_{1:t-1} ,\modpar) &= 
        \dens(\obs_{t}|\state_{t}, \modpar), \quad t=1,2, \ldots, \nT .
        \label{eq:hmmresponses}
\end{align} Making use of these conditional independencies, the joint
distribution of observations and states can be stated as
\begin{equation}
\dens(\obs_{1:\nT}, \state_{1:\nT}|\modpar) = \Prob(\state_1|\modpar) \dens(\obs_1 | \state_1, \modpar) \prod_{t=2}^{\nT} 
\Prob(\state_t|\state_{t-1},\modpar) \dens(\obs_t|\state_t,\modpar) .
\label{eq:gHMM_joint}
\end{equation} The likelihood function (i.e.~the marginal distribution
of the observations as a function of the model parameters) can then be
written as \begin{equation}
\label{eq:marginal_observation_dist}
L(\modpar|\obs_{1:T}) = 
\sum_{\fstate_{1:\nT} \in \mathcal{S}^T} \dens(\obs_{1:\nT}, \state_{1:\nT} = \fstate_{1:\nT} |\modpar) ,
\end{equation} where the summation is over all possible state sequences
(i.e.~\(\mathcal{S}^T\) is the set of all possible sequences of states).
Rather than actually summing over all possible state sequences, the
forward-backward algorithm \citep{Rabiner1989} is used to efficiently
calculate this likelihood. For more information on hidden Markov models,
see also \citet{visser2021hidden}.

\hypertarget{missing-data}{%
\subsection{Missing data}\label{missing-data}}

The canonical references for statistical inference with missing data are
\citet{rubin1976inference} and \citet{little2014statistical}. Here we
summarise the main ideas and results from those sources, as relevant to
the present topic.

Let \(\obs_{1:\nT}\), the sequence of all response variables, be
partitioned into a set of observed values,
\(\mathcal{\obs}_{\text{obs}} \subseteq \obs_{1:\nT}\), and a set of
missing values, \(\mathcal{\obs}_{\text{miss}} \subseteq \obs_{1:\nT}\).
Let \(M_{1:\nT}\) be vector of indicator variables with values
\(M_t = 1\) if \(\obs_t \in \mathcal{\obs}_{\text{miss}}\) (the
observation at time \(t\) is missing), and \(M_t = 0\) otherwise. In
addition to \(\modpar\), the parameters of the hidden Markov model for
the observed data \(\obs\), let \(\gvc{\phi}\) denote the parameter
vector of the statistical model of missingness (i.e.~the model of
\(M_{1:\nT}\)).

We can define the ``full'' likelihood function as \begin{equation}
\label{eq:joint_missing_likelihood}
L_\text{full}{(\modpar,\gvc{\phi}|\mathcal{\obs}_{\text{obs}},M_{1:\nT})} \propto \int \dens{(\mathcal{\obs}_{\text{obs}},\mathcal{\obs}_{\text{miss}}|\modpar)} \dens{(M_{1:\nT}|\mathcal{\obs}_{\text{obs}},\mathcal{\obs}_{\text{miss}},\gvc{\phi})} d \mathcal{\obs}_\text{miss} ,
\end{equation} that is, as any function proportional to
\(\dens(\mathcal{\obs}_{\text{obs}},M_{1:\nT}|\modpar,\gvc{\phi})\).
Note that this is a marginal density, hence the integration over all
possible values of the missing data. In this general case, we allow
missingness to depend on the ``complete'' data \(\obs_{1:\nT}\), so
including the missing values \(\mathcal{Y}_\text{miss}\) (for instance,
it might be the case that missing values occur when the true value of
\(\obs_t\) is relatively high).

Ignoring the missing data, the likelihood can be defined as
\begin{equation}
\label{eq:ignoring_missing_likelihood}
L_\text{ign}(\modpar|\mathcal{\obs}_{\text{obs}}) \propto \dens(\mathcal{\obs}_{\text{obs}}|\modpar) ,
\end{equation} that is, as any function proportional to
\(\dens(\mathcal{\obs}_\text{obs}|\modpar)\). An important question is
when inference for \(\modpar\) based on
(\ref{eq:joint_missing_likelihood}) and
(\ref{eq:ignoring_missing_likelihood}) give the same results. Note that
both likelihood functions need only be known up to a constant of
proportionality as only relative likelihoods need to be known for
maximizing the likelihood or computing likelihood ratio's. The question
is thus when (\ref{eq:ignoring_missing_likelihood}) is proportional to
(\ref{eq:joint_missing_likelihood}).

As shown by \citet{rubin1976inference}, inference on \(\modpar\) based
on (\ref{eq:joint_missing_likelihood}) and
(\ref{eq:ignoring_missing_likelihood}) will give identical results when
(1) \(\modpar\) and \(\gvc{\phi}\) are separable (i.e.~the joint
parameter space is the product of the parameter space for \(\modpar\)
and \(\gvc{\phi}\)), and (2) the following holds: \begin{equation}
\label{eq:MAR_definition}
\dens(M_{1:\nT}|\mathcal{\obs}_{\text{obs}},\mathcal{\obs}_{\text{miss}},\gvc{\phi}) = \dens(M_{1:\nT}|\mathcal{\obs}_{\text{obs}},\gvc{\phi}) \quad \quad \text{for all } \mathcal{\obs}_{\text{miss}}, \gvc{\phi}, 
\end{equation} i.e.~whether data is missing does not depend on the
missing values. In this case, data is said to be missing at random
(MAR), and the joint density can be factored as \begin{align*}
\dens{(\mathcal{\obs}_{\text{obs}},M_{1:\nT}|\modpar,\gvc{\phi})} &= \dens{(M_{1:\nT}|\mathcal{\obs}_{\text{obs}},\gvc{\phi})} \times \int \dens{(\mathcal{\obs}_{\text{obs}},\mathcal{\obs}_{\text{miss}}|\modpar)}  d \mathcal{\obs}_\text{miss} \\
&= \dens{(M_{1:\nT}|\mathcal{\obs}_{\text{obs}},\gvc{\phi})} \times \dens(\mathcal{\obs}_{\text{obs}}|\modpar) ,
\end{align*} which indicates that, as a function of \(\modpar\),
\(L_\text{full}(\modpar,\gvc{\phi}|\mathcal{\obs}_\text{obs},M_{1:\nT}) \propto L_\text{ign}(\modpar|\mathcal{\obs}_\text{obs})\).
Hence, when data is MAR, the missing data, and the mechanism leading to
it, can be ignored in inference for \(\modpar\). A special case of MAR
is data which is ``missing completely at random'' (MCAR), where
\begin{equation}
\label{eq:MCAR_definition}
\dens(M_{1:\nT}|\mathcal{\obs}_{\text{obs}},\mathcal{\obs}_{\text{miss}},\gvc{\phi}) = \dens(M_{1:\nT}|\gvc{\phi}) .
\end{equation}

When the equality in (\ref{eq:MAR_definition}) does not hold, data is
said to be missing not at random (MNAR). In this case, ignoring the
missing data will generally lead to biased parameter estimates of
\(\theta\). Valid inference of \(\theta\) requires working with the full
likelihood function of (\ref{eq:joint_missing_likelihood}), so
explicitly accounting for missingness.

\hypertarget{missing-data-in-hidden-markov-models}{%
\subsection{Missing data in hidden Markov
models}\label{missing-data-in-hidden-markov-models}}

Hidden Markov models by definition include missing data, as the hidden
states are unobservable (i.e.~always missing). When there are no missing
values for the observed variable \(\obs\), it is easy to see that
inference on \(\modpar\) in HMMs targets the correct likelihood.
Replacing \(\mathcal{\obs}_{\text{miss}}\) by \(\state_{1:\nT}\), and
noting that
\(\Prob(M_{1:\nT}|\mathcal{\obs}_{\text{obs}},\state_{1:\nT}) = \Prob(M_{1:\nT}) = 1\),
the missing states can be considered missing completely at random
(MCAR).

We will now focus on the case where the observable response variable
\(\obs\) does have missing values. The full likelihood, which also
depends on the hidden states, can be defined as \begin{equation}
L_\text{full}(\modpar,\gvc{\phi}|\mathcal{\obs}_{\text{obs}},M_{1:\nT}) \propto \sum_{\fstate_{1:\nT} \in \mathcal{S}^\nT} \int \dens{(\mathcal{\obs}_{\text{obs}},\mathcal{\obs}_{\text{miss}}, \fstate_{1:\nT} |\modpar)} \dens{(M_{1:\nT}|\mathcal{\obs}_{\text{obs}},\mathcal{\obs}_{\text{miss}},\fstate_{1:\nT},\gvc{\phi})} d \mathcal{\obs}_\text{miss} ,
\end{equation} while the likelihood ignoring missing data as
\begin{equation}
L_\text{ign}(\modpar|\mathcal{\obs}_{\text{obs}}) \propto \sum_{\fstate_{1:\nT} \in \mathcal{S}^\nT}  \dens{(\mathcal{\obs}_{\text{obs}}, \fstate_{1:\nT} |\modpar)} .
\end{equation}

\hypertarget{missing-at-random-mar}{%
\subsubsection{Missing at random (MAR)}\label{missing-at-random-mar}}

When the data is missing at random (\ref{eq:MAR_definition}), then
\begin{align}
\notag L_\text{full}(\modpar,\gvc{\phi}|\mathcal{\obs}_{\text{obs}},M_{1:\nT}) &\propto
\sum_{\fstate_{1:\nT} \in \mathcal{S}^\nT} \int \dens{(\mathcal{\obs}_{\text{obs}},\mathcal{\obs}_{\text{miss}}, \fstate_{1:\nT} |\modpar)} \dens{(M_{1:\nT}|\mathcal{\obs}_{\text{obs}},\gvc{\phi})} d \mathcal{\obs}_\text{miss} \\
\label{eq:HMM_mss_lik} &= \dens{(M_{1:\nT}|\mathcal{\obs}_{\text{obs}}, \gvc{\phi})} \times \left( \sum_{\fstate_{1:\nT} \in \mathcal{S}^\nT} \int \dens{(\mathcal{\obs}_{\text{obs}},\mathcal{\obs}_{\text{miss}}, \fstate_{1:\nT} |\modpar)} d \mathcal{\obs}_\text{miss} \right)
\end{align} and hence missingness is ignorable in inference of
\(\modpar\). Furthermore, defining \begin{align}
\notag \dens^*(\obs_t|\state_t,\modpar) &= \mathbb{I}_{\obs_t \in \mathcal{\obs}_{\text{obs}}} \dens(\obs_t | \state_t, \modpar) + \mathbb{I}_{\obs_t \in \mathcal{\obs}_{\text{miss}}} \int \dens(\obs_t | \state_t, \modpar) d \obs_t \\ 
&= \mathbb{I}_{\obs_t \in \mathcal{\obs}_{\text{obs}}} \dens(\obs_t | \state_t, \modpar) + \mathbb{I}_{\obs_t \in \mathcal{\obs}_{\text{miss}}} \times 1 ,
\end{align} where the indicator variable \(\mathbb{I}_x = 1\) if
condition \(x\) is true and 0 otherwise, we can write the part of the
full likelihood (\ref{eq:HMM_mss_lik}) relevant to inference on
\(\modpar\) as \begin{align*}
\sum_{\fstate_{1:\nT} \in \mathcal{S}^\nT} \int \dens( \mathcal{\obs}_{\text{obs}},\mathcal{\obs}_{\text{miss}}, \fstate_{1:\nT} | \modpar) d \mathcal{\obs}_{\text{miss}} &= \Prob(\state_1|\modpar) \dens^*(\obs_1 | \state_1, \modpar)  \prod_{t=2}^{\nT} 
\Prob(\state_t|\state_{t-1},\modpar) \dens^*(\obs_t|\state_t,\modpar) ,
\end{align*} which shows that a principled way to deal with missing
observations is to set \(\dens(\obs_t|\state_t) = 1\) for all
\(\obs_t \in \mathcal{\obs}_\text{miss}\). Note that it is necessary to
include time points with missing observations in this way to allow the
state probabilities to be computed properly. While this result is known
\citep[e.g.][]{zucchini2017hidden}, we have not come across its
derivation in the form above.

\hypertarget{state-dependent-missingness-mnar}{%
\subsubsection{State-dependent missingness
(MNAR)}\label{state-dependent-missingness-mnar}}

If data is not MAR, there is some dependence between whether
observations are missing or not, and the true unobserved values. There
are many forms this dependence can take, and modelling the dependence
accurately may require substantial knowledge of the domain to which the
data applies. Here, we take a pragmatic approach, and model this
dependence through the hidden states. That is, we assume \(M\) and
\(\obs\) are conditionally independent, given the hidden states:
\begin{equation*}
\dens{(M_t, \obs_t|\state_t)} = \dens{(M_t|\state_t)} \dens{(\obs_t|\state_t)} .
\end{equation*} This is not an overly restrictive assumption, as the
number of hidden states can be chosen to allow for intricate patterns of
(marginal) dependence between \(M\) and \(\obs\) at a single time point,
as well as over time. For example, increased probability of missingness
for high values of \(\obs\) can be captured through a state which is
simultaneously associated with high values of \(\obs\) and a high
probability of \(M=1\). A high probability of a missing observation at
\(t+1\) \emph{after} a high (observed) value of \(\obs_t\) can be
captured with a state \(s\) associated with high values of \(\obs\), a
state \(s' \neq s\) associated with a high probability of \(M=1\), and a
high transition probability \(P(S_{t+1} = s'|S_{t} = s)\).

Under the assumption that missingness depends only on the hidden states:
\begin{equation*}
\dens{(M_{1:\nT}|\mathcal{\obs}_{\text{obs}},\mathcal{\obs}_{\text{miss}},\state_{1:\nT},\gvc{\phi})} = \dens{(M_{1:\nT}|\state_{1:\nT},\gvc{\phi})} ,
\end{equation*} the full likelihood becomes \begin{align*}
L_\text{full}(\modpar,\gvc{\phi}|\mathcal{\obs}_{\text{obs}},M_{1:\nT}) &\propto \sum_{\fstate_{1:\nT} \in \mathcal{S}^\nT} \int \dens{(\mathcal{\obs}_{\text{obs}},\mathcal{\obs}_{\text{miss}}, \fstate_{1:\nT} |\modpar)} \dens{(M_{1:\nT}|\mathcal{\obs}_{\text{obs}},\mathcal{\obs}_{\text{miss}},\fstate_{1:\nT},\gvc{\phi})} d \mathcal{\obs}_\text{miss} \\
&= \sum_{\fstate_{1:\nT} \in \mathcal{S}^\nT}  \dens{(M_{1:\nT}|\fstate_{1:\nT},\gvc{\phi})} \times \int \dens{(\mathcal{\obs}_{\text{obs}},\mathcal{\obs}_{\text{miss}}, \fstate_{1:\nT} |\modpar)} d \mathcal{\obs}_\text{miss} \\
&= \sum_{\fstate_{1:\nT} \in \mathcal{S}^\nT}  \dens{(M_{1:\nT}|\fstate_{1:\nT},\gvc{\phi})} \times \dens{(\mathcal{\obs}_{\text{obs}}, \fstate_{1:\nT} |\modpar)} .
\end{align*} This shows that, although \(M\) does not directly depend on
\(\mathcal{\obs}_\text{miss}\), because both \(M\) and \(Y\) depend on
\(\state\), the role of the \(\dens{(M|\state,\gvc{\phi})}\) term is
more than a scaling factor in the likelihood, and hence missingness is
not ignorable.

\hypertarget{overview}{%
\subsection{Overview}\label{overview}}

When data is MNAR and missingness is not ignorable, valid inference on
\(\modpar\) requires including a submodel for \(M\) in the overall
model. That is, the HMM should be for multivariate data \(\obs\) and
\(M\). The objective of the present paper is to show the potential
benefits of including a relatively simple model for \(M\) in hidden
Markov models, by assuming missingness is state-dependent. We provide
results from a simulation study, as well as an example with real data.
The simulations assess the accuracy of parameter estimates and state
recovery in situations where missingness is MAR or MNAR and dependent on
the hidden state, in situations where the state-conditional
distributions of the observations are relatively well separated or more
overlapping. We then discuss a situation where missingness is
time-dependent (but not state-dependent). This is a situation where
missingness is in fact MCAR, and where a misspecified model which
assumes missingness is state-dependent may lead to biased results.
Finally, we apply the models to a real data example, involving a
clinical trial comparing the effect of real and placebo medication on
the severity of schizophrenic symptoms.

\hypertarget{simulation-study}{%
\section{Simulation study}\label{simulation-study}}

\begin{figure}
\includegraphics[width=\linewidth]{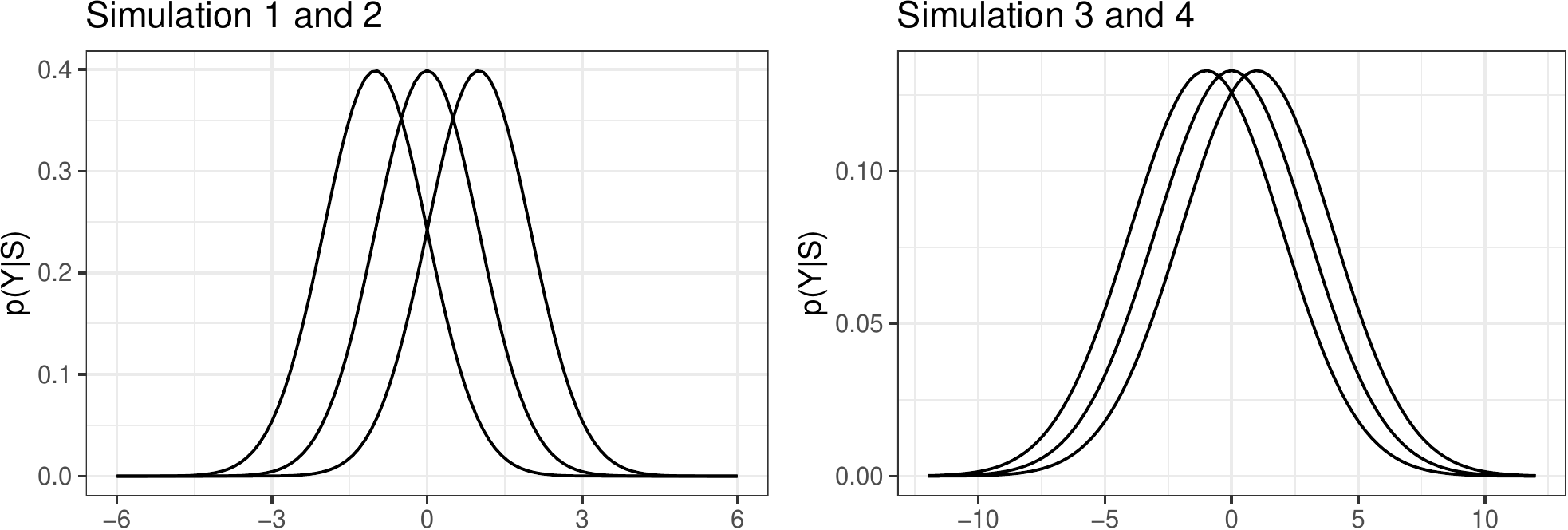} \caption{State-conditional response distributions in the simulation studies. In Simulation 1 and 2, states are reasonably well-separated, although there is still considerable overlap of the distributions. In Simulation 3 and 4, states less well-separated.}\label{fig:simulation-response-distributions}
\end{figure}

To assess the potential benefits of including a state-dependent
missingness model in a HMM, we conducted a simulation study, focussing
on a three-state hidden Markov model with a univariate Normal
distributed response variable\footnote{All code for the simulations, and
  the analysis of the application, is available at
  \url{https://github.com/depmix/hmm-missing-data-paper}.}. We simulated
four scenario's. In Simulation 1 and 2 (Figure
\ref{fig:simulation-response-distributions}), the states are reasonably
well-separated, with means \(\mu_1 = -1\), \(\mu_2 = 0\), \(\mu_3 = 1\)
and standard deviations \(\sigma_1 = \sigma_2 = \sigma_3 = 1\). Note
that there is still considerable overlap in the state-conditional
response distributions, as would be expected in many real applications
of HMMs. In Simulation 1, missingness was state-dependent (i.e.~MNAR),
with \(\Prob(M_t = 1|S_t = 1) = .05\), \(\Prob(M_t = 1|S_t = 2) = .25\),
and \(\Prob(M_t = 1|S_t = 2) = .5\). In Simulation 2, missingness was
independent of the state (MAR), with
\(\Prob(M_t = 1|S_t = i) = \Prob(M_t = 1) = .25\). In Simulation 3 and 4
(Figure \ref{fig:simulation-response-distributions}), the states were
rather less well-separated, with means as for Simulation 1 and 2, but
standard deviations \(\sigma_i = 3\). Here, the overlap of the
state-conditional response distributions is much higher than in
Simulation 1 and 2, and identification of the hidden states will be more
difficult. In Simulation 3, missingness was state-dependent (MNAR) in
the same manner as Simulation 1, while in Simulation 4, missingness was
state-independent (MAR) as for Simulation 2. In all simulations, the
initial state probabilities were \(\pi_1 = \Prob(S_1 = 1) = .8\),
\(\pi_2 = \pi_3 = .1\), and the state-transition matrix was
\begin{equation*}
\mat{A} = \left[ \begin{matrix} .75 &  .125 & .125 \\ .125 & .75 & .125 \\ .125 & .125 & .75
\end{matrix} \right] .
\end{equation*} In each simulation, we simulated a total of 1000 data
sets, each consisting of \(\no = 100\) replications of a time-series of
length \(\nT = 50\). We denote observations in such replicated time
series as \(\obs_{i,t}\), with \(i=1,\ldots,\no\) and
\(t = 1, \ldots, \nT\). Data was generated according to a 3-state hidden
Markov model. For MAR cases, the non-missing observations are
distributed as \begin{equation}
\label{eq:MAR-distribution}
\dens(\obs_{i,t}|\state_{i,t} = j) = \mathbf{Normal}(\obs_{i,t}|\mu_j,\sigma_j) .
\end{equation} In the MNAR cases, the missingness variable \(M\) and the
response variable \(\obs\) were conditionally independent given the
hidden state: \begin{equation}
\label{eq:MNAR-distribution}
\dens(\obs_{i,t},M_{i,t}|\state_{i,t} = j) = \mathbf{Bernouilli}(M_{i,t}|\phi_j) \times \mathbf{Normal}(\obs_{i,t}|\mu_j,\sigma_j)
\end{equation} Data sets were simulated by first generating the hidden
state sequences \(\state_{i,1:\nT}\) according to the initial state and
transition probabilities. Then, the observations \(\obs_{i,1:\nT}\) were
sampled according to the state-conditional distributions
\(\dens(\obs_{i,t}|\state_{i,t})\). Finally, random observations were
set to missing values according to the missingness distributions
\(\Prob(M_{i,t}|\state_{i,t})\).

We fitted two 3-state hidden Markov models to each data-set. In the MAR
models, observed responses were assumed to be distributed according to
(\ref{eq:MAR-distribution}), and in the MNAR models, the observed
responses and missingness indicators were assumed to be distributed
according to (\ref{eq:MNAR-distribution}). Parameters were estimated by
maximum likelihood, using the Expectation-Maximisation algorithm, as
implemented in depmixS4 \citep{depmixS4}. To speed up convergence,
starting values were set to the true parameter values. Although such
initialization is obviously not possible in real applications, we are
interested in the quality of parameter estimates at the global maximum
likelihood solution, and setting starting values to the true parameters
makes it more likely to arrive at the global maximum. In real
applications, one would need to use a sufficient number of randomly
generated starting values to find the global maximum.

\begin{table}

\caption{\label{tab:table-simulation-1}Results of Simulation 1 (MNAR, low variance). Values shown are the true value of each parameter, and the mean (mean), standard deviation (SD), and mean absolute error (MAE) of the parameter estimates, for both the MAR and MNAR model. The value of "rel. MAE" is the ratio of the mean absolute error of the MAR over the MNAR model.}
\centering
\begin{tabular}[t]{lrrrrrrrr}
\toprule
\multicolumn{1}{c}{ } & \multicolumn{1}{c}{ } & \multicolumn{3}{c}{MAR} & \multicolumn{3}{c}{MNAR} & \multicolumn{1}{c}{ } \\
\cmidrule(l{3pt}r{3pt}){3-5} \cmidrule(l{3pt}r{3pt}){6-8}
parameter & true value & mean & SD & MAE & mean & SD & MAE & rel. MAE\\
\midrule
$\mu_1$ & -1.000 & -1.010 & 0.131 & 0.094 & -1.017 & 0.097 & 0.072 & 0.767\\
$\mu_2$ & 0.000 & 0.015 & 0.277 & 0.223 & 0.014 & 0.228 & 0.180 & 0.807\\
$\mu_3$ & 1.000 & 1.113 & 0.286 & 0.231 & 1.053 & 0.252 & 0.186 & 0.803\\
\addlinespace
$\sigma_1$ & 1.000 & 0.998 & 0.051 & 0.033 & 0.995 & 0.034 & 0.026 & 0.785\\
$\sigma_2$ & 1.000 & 0.972 & 0.111 & 0.079 & 0.979 & 0.085 & 0.064 & 0.809\\
$\sigma_3$ & 1.000 & 0.959 & 0.104 & 0.077 & 0.979 & 0.085 & 0.061 & 0.801\\
\addlinespace
$\pi_1$ & 0.800 & 0.834 & 0.146 & 0.117 & 0.776 & 0.110 & 0.083 & 0.715\\
$\pi_2$ & 0.100 & 0.118 & 0.152 & 0.111 & 0.131 & 0.130 & 0.105 & 0.944\\
$\pi_3$ & 0.100 & 0.049 & 0.048 & 0.062 & 0.093 & 0.066 & 0.055 & 0.890\\
\addlinespace
$a_{11}$ & 0.750 & 0.774 & 0.080 & 0.064 & 0.743 & 0.055 & 0.039 & 0.613\\
$a_{12}$ & 0.125 & 0.144 & 0.094 & 0.068 & 0.139 & 0.077 & 0.061 & 0.896\\
$a_{13}$ & 0.125 & 0.082 & 0.055 & 0.057 & 0.118 & 0.054 & 0.044 & 0.765\\
\addlinespace
$a_{21}$ & 0.125 & 0.144 & 0.086 & 0.065 & 0.124 & 0.062 & 0.048 & 0.729\\
$a_{22}$ & 0.750 & 0.759 & 0.116 & 0.087 & 0.754 & 0.096 & 0.070 & 0.812\\
$a_{23}$ & 0.125 & 0.097 & 0.082 & 0.068 & 0.122 & 0.075 & 0.058 & 0.850\\
\addlinespace
$a_{31}$ & 0.125 & 0.146 & 0.085 & 0.068 & 0.118 & 0.050 & 0.039 & 0.579\\
$a_{32}$ & 0.125 & 0.166 & 0.128 & 0.103 & 0.138 & 0.092 & 0.070 & 0.679\\
$a_{33}$ & 0.750 & 0.688 & 0.111 & 0.090 & 0.744 & 0.076 & 0.056 & 0.623\\
\addlinespace
$p(M=1|S=1)$ & 0.050 & - & - & - & 0.048 & 0.021 & 0.017 & -\\
$p(M=1|S=2)$ & 0.250 & - & - & - & 0.247 & 0.073 & 0.057 & -\\
$p(M=1|S=3)$ & 0.500 & - & - & - & 0.507 & 0.058 & 0.040 & -\\
\bottomrule
\end{tabular}
\end{table}

\begin{table}

\caption{\label{tab:table-simulation-2}Results of Simulation 2 (MAR, low variance). Values shown are the true value of each parameter, and the mean (mean), standard deviation (SD), and mean absolute error (MAE) of the parameter estimates, for both the MAR and MNAR model. The value of "rel. MAE" is the ratio of the mean absolute error of the MAR over the MNAR model.}
\centering
\begin{tabular}[t]{lrrrrrrrr}
\toprule
\multicolumn{1}{c}{ } & \multicolumn{1}{c}{ } & \multicolumn{3}{c}{MAR} & \multicolumn{3}{c}{MNAR} & \multicolumn{1}{c}{ } \\
\cmidrule(l{3pt}r{3pt}){3-5} \cmidrule(l{3pt}r{3pt}){6-8}
parameter & true value & mean & SD & MAE & mean & SD & MAE & rel. MAE\\
\midrule
$\mu_1$ & -1.000 & -1.061 & 0.194 & 0.132 & -1.062 & 0.200 & 0.134 & 1.014\\
$\mu_2$ & 0.000 & -0.019 & 0.287 & 0.229 & -0.022 & 0.288 & 0.230 & 1.005\\
$\mu_3$ & 1.000 & 1.048 & 0.213 & 0.158 & 1.038 & 0.213 & 0.157 & 0.991\\
\addlinespace
$\sigma_1$ & 1.000 & 0.978 & 0.067 & 0.047 & 0.978 & 0.070 & 0.047 & 1.000\\
$\sigma_2$ & 1.000 & 0.969 & 0.105 & 0.079 & 0.969 & 0.107 & 0.081 & 1.021\\
$\sigma_3$ & 1.000 & 0.981 & 0.070 & 0.050 & 0.982 & 0.071 & 0.049 & 0.993\\
\addlinespace
$\pi_1$ & 0.800 & 0.739 & 0.177 & 0.130 & 0.737 & 0.178 & 0.132 & 1.015\\
$\pi_2$ & 0.100 & 0.169 & 0.187 & 0.144 & 0.171 & 0.187 & 0.145 & 1.012\\
$\pi_3$ & 0.100 & 0.092 & 0.070 & 0.057 & 0.092 & 0.069 & 0.057 & 0.995\\
\addlinespace
$a_{11}$ & 0.750 & 0.727 & 0.102 & 0.064 & 0.726 & 0.102 & 0.065 & 1.006\\
$a_{12}$ & 0.125 & 0.155 & 0.112 & 0.082 & 0.156 & 0.115 & 0.083 & 1.020\\
$a_{13}$ & 0.125 & 0.118 & 0.066 & 0.051 & 0.118 & 0.062 & 0.050 & 0.975\\
\addlinespace
$a_{21}$ & 0.125 & 0.125 & 0.080 & 0.061 & 0.127 & 0.083 & 0.062 & 1.013\\
$a_{22}$ & 0.750 & 0.751 & 0.112 & 0.084 & 0.749 & 0.116 & 0.086 & 1.025\\
$a_{23}$ & 0.125 & 0.125 & 0.081 & 0.061 & 0.125 & 0.083 & 0.063 & 1.034\\
\addlinespace
$a_{31}$ & 0.125 & 0.112 & 0.063 & 0.051 & 0.112 & 0.062 & 0.049 & 0.973\\
$a_{32}$ & 0.125 & 0.153 & 0.108 & 0.082 & 0.150 & 0.107 & 0.081 & 0.982\\
$a_{33}$ & 0.750 & 0.735 & 0.096 & 0.067 & 0.738 & 0.095 & 0.066 & 0.984\\
\addlinespace
$p(M=1|S=1)$ & 0.250 & - & - & - & 0.250 & 0.046 & 0.027 & -\\
$p(M=1|S=2)$ & 0.250 & - & - & - & 0.248 & 0.056 & 0.038 & -\\
$p(M=1|S=3)$ & 0.250 & - & - & - & 0.247 & 0.046 & 0.028 & -\\
\bottomrule
\end{tabular}
\end{table}

\begin{table}

\caption{\label{tab:table-simulation-3}Results of Simulation 3 (MNAR, high variance). Values shown are the true value of each parameter, and the mean (mean), standard deviation (SD), and mean absolute error (MAE) of the parameter estimates, for both the MAR and MNAR model. The value of "rel. MAE" is the ratio of the mean absolute error of the MAR over the MNAR model.}
\centering
\begin{tabular}[t]{lrrrrrrrr}
\toprule
\multicolumn{1}{c}{ } & \multicolumn{1}{c}{ } & \multicolumn{3}{c}{MAR} & \multicolumn{3}{c}{MNAR} & \multicolumn{1}{c}{ } \\
\cmidrule(l{3pt}r{3pt}){3-5} \cmidrule(l{3pt}r{3pt}){6-8}
parameter & true value & mean & SD & MAE & mean & SD & MAE & rel. MAE\\
\midrule
$\mu_1$ & -1.000 & -1.663 & 0.932 & 0.761 & -1.198 & 0.628 & 0.315 & 0.414\\
$\mu_2$ & 0.000 & -0.314 & 0.484 & 0.461 & -0.110 & 0.470 & 0.409 & 0.888\\
$\mu_3$ & 1.000 & 1.480 & 1.214 & 0.923 & 1.383 & 0.956 & 0.609 & 0.661\\
\addlinespace
$\sigma_1$ & 3.000 & 2.765 & 0.459 & 0.347 & 2.911 & 0.330 & 0.189 & 0.543\\
$\sigma_2$ & 3.000 & 2.889 & 0.455 & 0.302 & 2.967 & 0.333 & 0.215 & 0.713\\
$\sigma_3$ & 3.000 & 2.703 & 0.512 & 0.406 & 2.773 & 0.479 & 0.326 & 0.803\\
\addlinespace
$\pi_1$ & 0.800 & 0.546 & 0.362 & 0.355 & 0.657 & 0.281 & 0.217 & 0.611\\
$\pi_2$ & 0.100 & 0.346 & 0.380 & 0.333 & 0.253 & 0.291 & 0.231 & 0.694\\
$\pi_3$ & 0.100 & 0.108 & 0.174 & 0.129 & 0.090 & 0.091 & 0.077 & 0.601\\
\addlinespace
$a_{11}$ & 0.750 & 0.651 & 0.231 & 0.183 & 0.712 & 0.153 & 0.099 & 0.543\\
$a_{12}$ & 0.125 & 0.190 & 0.215 & 0.160 & 0.144 & 0.145 & 0.105 & 0.660\\
$a_{13}$ & 0.125 & 0.159 & 0.186 & 0.139 & 0.144 & 0.124 & 0.091 & 0.659\\
\addlinespace
$a_{21}$ & 0.125 & 0.106 & 0.172 & 0.124 & 0.106 & 0.108 & 0.085 & 0.687\\
$a_{22}$ & 0.750 & 0.787 & 0.232 & 0.185 & 0.784 & 0.135 & 0.109 & 0.590\\
$a_{23}$ & 0.125 & 0.107 & 0.158 & 0.115 & 0.110 & 0.105 & 0.085 & 0.742\\
\addlinespace
$a_{31}$ & 0.125 & 0.152 & 0.183 & 0.136 & 0.131 & 0.126 & 0.096 & 0.704\\
$a_{32}$ & 0.125 & 0.166 & 0.199 & 0.143 & 0.145 & 0.141 & 0.105 & 0.738\\
$a_{33}$ & 0.750 & 0.682 & 0.234 & 0.184 & 0.724 & 0.151 & 0.108 & 0.587\\
\addlinespace
$p(M=1|S=1)$ & 0.050 & - & - & - & 0.076 & 0.122 & 0.059 & -\\
$p(M=1|S=2)$ & 0.250 & - & - & - & 0.241 & 0.155 & 0.126 & -\\
$p(M=1|S=3)$ & 0.500 & - & - & - & 0.489 & 0.134 & 0.092 & -\\
\bottomrule
\end{tabular}
\end{table}

\begin{table}

\caption{\label{tab:table-simulation-4}Results of Simulation 4 (MAR, high variance). Values shown are the true value of each parameter, and the mean (mean), standard deviation (SD), and mean absolute error (MAE) of the parameter estimates, for both the MAR and MNAR model. The value of "rel. MAE" is the ratio of the mean absolute error of the MAR over the MNAR model.}
\centering
\begin{tabular}[t]{lrrrrrrrr}
\toprule
\multicolumn{1}{c}{ } & \multicolumn{1}{c}{ } & \multicolumn{3}{c}{MAR} & \multicolumn{3}{c}{MNAR} & \multicolumn{1}{c}{ } \\
\cmidrule(l{3pt}r{3pt}){3-5} \cmidrule(l{3pt}r{3pt}){6-8}
parameter & true value & mean & SD & MAE & mean & SD & MAE & rel. MAE\\
\midrule
$\mu_1$ & -1.000 & -1.650 & 1.002 & 0.801 & -1.658 & 1.107 & 0.815 & 1.018\\
$\mu_2$ & 0.000 & -0.171 & 0.539 & 0.432 & -0.178 & 0.542 & 0.437 & 1.010\\
$\mu_3$ & 1.000 & 1.468 & 1.063 & 0.778 & 1.473 & 1.070 & 0.788 & 1.014\\
\addlinespace
$\sigma_1$ & 3.000 & 2.719 & 0.468 & 0.383 & 2.720 & 0.473 & 0.375 & 0.981\\
$\sigma_2$ & 3.000 & 2.911 & 0.441 & 0.299 & 2.918 & 0.412 & 0.279 & 0.934\\
$\sigma_3$ & 3.000 & 2.728 & 0.504 & 0.377 & 2.732 & 0.478 & 0.365 & 0.968\\
\addlinespace
$\pi_1$ & 0.800 & 0.528 & 0.345 & 0.352 & 0.522 & 0.338 & 0.357 & 1.012\\
$\pi_2$ & 0.100 & 0.330 & 0.367 & 0.316 & 0.344 & 0.359 & 0.320 & 1.014\\
$\pi_3$ & 0.100 & 0.141 & 0.199 & 0.149 & 0.134 & 0.192 & 0.142 & 0.951\\
\addlinespace
$a_{11}$ & 0.750 & 0.638 & 0.230 & 0.183 & 0.645 & 0.220 & 0.177 & 0.968\\
$a_{12}$ & 0.125 & 0.188 & 0.212 & 0.155 & 0.182 & 0.211 & 0.155 & 0.998\\
$a_{13}$ & 0.125 & 0.174 & 0.188 & 0.139 & 0.174 & 0.178 & 0.134 & 0.963\\
\addlinespace
$a_{21}$ & 0.125 & 0.111 & 0.166 & 0.121 & 0.103 & 0.157 & 0.119 & 0.984\\
$a_{22}$ & 0.750 & 0.774 & 0.223 & 0.175 & 0.787 & 0.209 & 0.170 & 0.972\\
$a_{23}$ & 0.125 & 0.114 & 0.152 & 0.111 & 0.110 & 0.139 & 0.110 & 0.986\\
\addlinespace
$a_{31}$ & 0.125 & 0.133 & 0.169 & 0.125 & 0.137 & 0.167 & 0.124 & 0.997\\
$a_{32}$ & 0.125 & 0.167 & 0.193 & 0.138 & 0.162 & 0.186 & 0.139 & 1.007\\
$a_{33}$ & 0.750 & 0.700 & 0.232 & 0.177 & 0.701 & 0.224 & 0.176 & 0.992\\
\addlinespace
$p(M=1|S=1)$ & 0.250 & - & - & - & 0.253 & 0.122 & 0.080 & -\\
$p(M=1|S=2)$ & 0.250 & - & - & - & 0.237 & 0.086 & 0.056 & -\\
$p(M=1|S=3)$ & 0.250 & - & - & - & 0.257 & 0.130 & 0.082 & -\\
\bottomrule
\end{tabular}
\end{table}

The results of simulation 1 (Table \ref{tab:table-simulation-1}) show
that, when states are relatively well separated, both models provide
parameter estimates which are, on average, reasonably close to the true
values. Both models have the tendency to estimate the means as more
dispersed, and the standard deviations as slightly smaller, then they
really are. While wrongly assuming MAR may not lead to overly biased
estimates, we see that the mean absolute error (MAE) for the MNAR model
is always smaller than that of the MAR model, reducing the estimation
error to as much as 58\%. As such, accounting for state-dependent
missingness increases the accuracy of the parameter estimates. We next
consider recovery of the hidden states, by comparing the true hidden
state sequences to the maximum a posteriori state sequences determined
by the Viterbi algorithm \citep[see][]{Rabiner1989, visser2021hidden}.
The MAR model recovers 53.13\% of the states, while the MNAR model
recovers 62.86\% of the states. The accuracy in recovering the hidden
states is thus higher in the model which correctly accounts for
missingness. Whilst the performance of neither model may seem overly
impressive, we should note that recovering the hidden states is a
non-trivial task when the state-conditional response distributions have
considerable overlap (see Figure
\ref{fig:simulation-response-distributions}) and states do not persist
for long periods of time (here, the true self-transitions probabilities
are \(a_{ii} = .75\), meaning that states have an average run-length of
4 consecutive time-points). When ignoring time-dependencies and treating
the observed data as coming from a bivariate mixture distribution over
\(\obs\) and \(M\), the maximum accuracy in classification would be
50.09\% for this data. The theoretical maximum classification accuracy
for the hidden Markov model is more difficult to establish, but
simulations show that the MNAR model with the true parameters recovers
66.51\% of the true states. For the MAR model, the approximate maximum
classification accuracy is 58.06\%.

The results of Simulation 2 (Table \ref{tab:table-simulation-2}) show
that when data is in fact MAR, both models provide roughly equally
accurate parameter estimates. While the MNAR model does not provide
better parameter estimates, including a model component for
state-dependent missingness does not seem to bias parameter estimates
compared to the MAR model. As can be seen, the state-wise missingness
probabilities are, on average, close to the true values of .25. Over all
parameters, the relative MAE of the models is 1.003 on average, which
shows the models perform equally well. In terms of recovering the hidden
states, the MAR model recovers 55.6\% of the states, while the MNAR
model recovers 55.63\% of the states. The somewhat reduced recovery rate
of the MNAR model compared to Simulation 1 is likely due to the fact
that here, missingness provides no information about the identity of the
hidden state. Here, the maximum classification accuracy is 42.91\% for a
mixture model, and approximately 60.45\% for the hidden Markov models.

In Simulation 3 (Table \ref{tab:table-simulation-3}) and 4 (Table
\ref{tab:table-simulation-4}) the states are less well-separated, making
accurate parameter estimation more difficult. Here, the tendency to
estimate the means as more dispersed and the standard deviations as
smaller than they are becomes more pronounced. For both models the
estimation error in Simulation 3 (Table \ref{tab:table-simulation-3}) is
larger than for Simulation 1, but comparing the MAE for both models
again shows the substantial benefits for including a missingness model.
Over all parameters, the relative MAE of the models is 0.658 on average,
which shows the MNAR model clearly outperforms the MAR model. In terms
of recovering the hidden states, the MAR model recovers 34.97\% of the
states, while the MNAR model recovers 45.27\% of the states. As in
Simulation 1, the MNAR model performs better in state identification.
For both models, performance is lower than in Simulation 1, reflecting
the increased difficulty due to increased overlap of the
state-conditional response distributions (Figure
\ref{fig:simulation-response-distributions}). Indeed, the performance of
the MAR model is close to chance (random assignment of states would give
an expected accuracy of 33.33\%). The maximum classification accuracy is
44.03\% for a mixture model, and approximately 54.04\% for the MNAR and
41.42\% for the MAR hidden Markov models.

When missingness is ignorable (Simulation 4), Like in Simulation 2,
inclusion of a missingness component in the HMM does not increase any
bias in the parameter estimates. Over all parameters, the relative MAE
of the models is 0.987 on average, which shows the models perform
roughly equally well. The model which ignores missingness recovers
35.51\% of the states, while the model with a missingness component
recovers 35.5\% of the states. For comparison, the maximum accuracy is
36.64\% for a mixture model, and 42.51\% for the hidden Markov models.

Taken together, these simulation results show that if missingness is
state-dependent, there is a substantial benefit to including a
(relatively simple) model for missingness in the HMM. When missingness
is in fact ignorable, including a missingness model is superfluous, but
does not bias the results. Hence, there appears to be little risk
associated to including a missingness model into the HMM.

In a final simulation, we assessed the performance of the models when
missingness is \emph{time-dependent}, rather than state-dependent.
Attrition is a common occurrence in longitudinal studies, meaning that
the probability of missing data often increases with time. In this
simulation, the probability of missing data varied with time \(t\)
through a logistic regression model: \begin{equation}
\Prob(M_{i,t} = 1) = \frac{1}{1+\exp(-(0.125 \times t - 5))} .
\end{equation} Here, the probability of missing data is very small
(0.008) at time 1, but increasing to rather high (0.777) at time 50. The
other parameters were the same as in Simulation 1 and 2 (i.e., the
states were relatively well-separated). In a model that specifies
missingness as state-dependent, but not time-dependent, this could
potentially result in biased parameter estimates. For instance, the
increased probability of missingness over time may be accounted for by
estimating states to have a different probability of missingness, and
estimating prior and transition probabilities to allow states with a
higher probability of missingness to occur more frequently later in
time. In addition to the two hidden Markov models estimated before, we
also estimated a hidden Markov model with a state- and time-dependent
model for missingness: \begin{equation}
\Prob(M_{i,t} = 1|S_{i,t} = j) = \frac{1}{1+\exp(-(\beta_{0,j} + \beta_{\text{time},j} \times t))}
\end{equation} This model should be able to capture the true pattern of
missingness, whilst the MNAR model which only includes state-dependent
missingness would not.

\begin{table}

\caption{\label{tab:table-simulation-5}Results of Simulation 5 (time-dependent missingness, low variance). Values shown are the true value of each parameter, and the mean (mean), standard deviation (SD), and mean absolute error (MAE) of the parameter estimates, for the MAR, MNAR (state), and MNAR (time) model. The value of "rel. MAE 1" is the ratio of the mean absolute error of the MAR over the MNAR (state) model, and the value of "rel. MAE 2" is the ratio of the mean absolute error of the MAR over the MNAR (time) model.}
\centering
\resizebox{\linewidth}{!}{
\begin{tabular}[t]{lrrrrrrrrrrrr}
\toprule
\multicolumn{1}{c}{ } & \multicolumn{1}{c}{ } & \multicolumn{3}{c}{MAR} & \multicolumn{3}{c}{MNAR (state)} & \multicolumn{3}{c}{MNAR (time)} & \multicolumn{2}{c}{ } \\
\cmidrule(l{3pt}r{3pt}){3-5} \cmidrule(l{3pt}r{3pt}){6-8} \cmidrule(l{3pt}r{3pt}){9-11}
parameter & true value & mean & SD & MAE & mean & SD & MAE & mean & SD & MAE & rel. MAE 1 & rel. MAE 2\\
\midrule
$\mu_1$ & -1.000 & -1.038 & 0.173 & 0.116 & -0.825 & 0.076 & 0.176 & -1.031 & 0.175 & 0.120 & 1.520 & 1.037\\
$\mu_2$ & 0.000 & -0.015 & 0.272 & 0.215 & -0.040 & 0.067 & 0.062 & -0.020 & 0.294 & 0.233 & 0.287 & 1.086\\
$\mu_3$ & 1.000 & 1.033 & 0.195 & 0.143 & 0.757 & 0.084 & 0.243 & 1.029 & 0.207 & 0.153 & 1.698 & 1.068\\
\addlinespace
$\sigma_1$ & 1.000 & 0.985 & 0.059 & 0.042 & 1.026 & 0.035 & 0.035 & 0.987 & 0.058 & 0.042 & 0.832 & 0.997\\
$\sigma_2$ & 1.000 & 0.967 & 0.111 & 0.082 & 1.278 & 0.033 & 0.278 & 0.966 & 0.112 & 0.083 & 3.370 & 1.010\\
$\sigma_3$ & 1.000 & 0.981 & 0.072 & 0.049 & 1.037 & 0.037 & 0.043 & 0.984 & 0.067 & 0.048 & 0.884 & 0.990\\
\addlinespace
$\pi_1$ & 0.800 & 0.758 & 0.151 & 0.110 & 0.894 & 0.066 & 0.100 & 0.760 & 0.152 & 0.109 & 0.913 & 0.993\\
$\pi_2$ & 0.100 & 0.150 & 0.161 & 0.123 & 0.001 & 0.032 & 0.101 & 0.148 & 0.162 & 0.122 & 0.819 & 0.990\\
$\pi_3$ & 0.100 & 0.092 & 0.061 & 0.050 & 0.105 & 0.059 & 0.047 & 0.092 & 0.062 & 0.051 & 0.943 & 1.023\\
\addlinespace
$a_{11}$ & 0.750 & 0.734 & 0.087 & 0.056 & 0.794 & 0.025 & 0.045 & 0.734 & 0.090 & 0.059 & 0.806 & 1.050\\
$a_{12}$ & 0.125 & 0.146 & 0.099 & 0.073 & 0.021 & 0.008 & 0.104 & 0.146 & 0.106 & 0.075 & 1.438 & 1.036\\
$a_{13}$ & 0.125 & 0.119 & 0.058 & 0.047 & 0.185 & 0.024 & 0.060 & 0.119 & 0.059 & 0.048 & 1.269 & 1.021\\
\addlinespace
$a_{21}$ & 0.125 & 0.128 & 0.088 & 0.063 & 0.000 & 0.001 & 0.125 & 0.131 & 0.097 & 0.069 & 1.983 & 1.091\\
$a_{22}$ & 0.750 & 0.747 & 0.114 & 0.083 & 1.000 & 0.001 & 0.250 & 0.738 & 0.131 & 0.092 & 2.994 & 1.104\\
$a_{23}$ & 0.125 & 0.125 & 0.082 & 0.062 & 0.000 & 0.000 & 0.125 & 0.130 & 0.091 & 0.067 & 2.031 & 1.082\\
\addlinespace
$a_{31}$ & 0.125 & 0.116 & 0.062 & 0.048 & 0.150 & 0.027 & 0.030 & 0.115 & 0.063 & 0.049 & 0.624 & 1.032\\
$a_{32}$ & 0.125 & 0.143 & 0.101 & 0.076 & 0.045 & 0.008 & 0.080 & 0.146 & 0.106 & 0.081 & 1.052 & 1.056\\
$a_{33}$ & 0.750 & 0.742 & 0.087 & 0.062 & 0.805 & 0.025 & 0.056 & 0.740 & 0.093 & 0.068 & 0.899 & 1.095\\
\addlinespace
$\beta_{0,1}$ & -5.000 & - & - & - & - & - & - & -6.149 & 25.490 & 1.497 & - & -\\
$\beta_{0,2}$ & -5.000 & - & - & - & - & - & - & -7.153 & 40.582 & 2.739 & - & -\\
$\beta_{0,3}$ & -5.000 & - & - & - & - & - & - & -6.472 & 38.998 & 1.861 & - & -\\
\addlinespace
$\beta_{\text{time},1}$ & 0.125 & - & - & - & - & - & - & 0.154 & 0.605 & 0.039 & - & -\\
$\beta_{\text{time},2}$ & 0.125 & - & - & - & - & - & - & 0.184 & 1.102 & 0.075 & - & -\\
$\beta_{\text{time},3}$ & 0.125 & - & - & - & - & - & - & 0.188 & 1.815 & 0.074 & - & -\\
\addlinespace
$p(M=1|S=1)$ & - & - & - & - & 0.040 & 0.015 & 0.207 & - & - & - & - & -\\
$p(M=1|S=2)$ & - & - & - & - & 0.552 & 0.024 & 0.305 & - & - & - & - & -\\
$p(M=1|S=3)$ & - & - & - & - & 0.067 & 0.022 & 0.181 & - & - & - & - & -\\
\bottomrule
\end{tabular}}
\end{table}

The results (Table \ref{tab:table-simulation-5}) show that, compared to
the MAR model, the MNAR model which misspecifies missingness as
state-dependent is inferior, resulting in more biased parameter
estimates. Over all parameters, the relative MAE of these two models is
1.353 on average, indicating the MAR model outperforms the MNAR (state)
model. To account for the increase in missing values later in time, the
MNAR (state) model estimates the probability of missingness as highest
for state 2, which is estimated to have a mean of close to 0, but an
increased standard deviation to incorporate observations from the other
two states. To make state 2 more prevalent over time, transition
probabilities to state 2 are relatively low from state 1 and 2
(parameters \(a_{12}\) and \(a_{32}\) respectively), whilst
self-transitions (\(a_{22}\)) are close to 1 (meaning that once in state
2, the hidden state sequence is very likely to remain in that state. The
prevalence of state 2 is thus increasing over time, and as this state
has a higher probability of missingness, so is the prevalence of missing
values. The MNAR (time) model, which allows missingness to depend on
both the hidden states and time, performs only slightly worse than the
MAR model, with an average relative MAE over all parameters of this
model compared to the MAR of 1.042. However, the MNAR (time) model is
able to capture the pattern of attrition (increased missing data over
time), whilst the MAR model is not. As such, the MNAR (time) model may
be deemed preferable to the MAR model, insofar as one is interested in
more than modelling the responses \(\obs\). In terms of recovering the
hidden states, the MAR model recovers 55.67\% of the states, and the
MNAR (time) model recovers 55.42\% of the states. The misspecified MNAR
(state) model recovers 50\% of the states. The maximum classification
accuracy for this data is 42.95\% for a mixture model, and approximately
59.91\% for the hidden Markov models.

This final simulation shows that when modelling patterns of missing data
in hidden Markov models, care should be taken in how this is done. An
increase in missing data over time could be due to an underlying higher
prevalence of states which result in more missing data, and/or a
state-independent increase in missingness over time. In applications
where the true reason and pattern of missingness is unknown, it is then
advisable to start by allowing for both state- and time-dependent
missing data, selecting simpler options when this is warranted by the
data.

\hypertarget{application-severity-of-schizophrenia-in-a-clinical-trial}{%
\section{Application: Severity of schizophrenia in a clinical
trial}\label{application-severity-of-schizophrenia-in-a-clinical-trial}}

Here, we apply our hidden Markov model with state-dependent missingness
to data from the National Institute of Mental Health Schizophrenia
Collaborative Study. The study concerns the assessment of
treatment-related changes in overall severity of mental illness. In this
study, 437 patients diagnosed with schizophrenia were randomly assigned
to receive either placebo (108 patients) or a drug (329 patients)
treatment, and their severity of their illness was rated by a clinician
at baseline (week 0), and at subsequent 1 week intervals (weeks 1--6),
with week 1, 3, and 6 as the intended main follow-up measurements. This
data has been made publicly available by Don Hedeker\footnote{\url{https://hedeker.people.uic.edu/SCHIZREP.DAT.txt}.}
and has been analysed numerous times. In particular,
\citet{hedeker1997application} focused on pattern mixture methods to
deal with missing data. \citet{yeh2010estimating} and
\citet{yeh2012intermittent} applied Markov and hidden Markov models,
respectively, assuming ratings were MAR.

\begin{figure}

{\centering \includegraphics[width=0.8\linewidth]{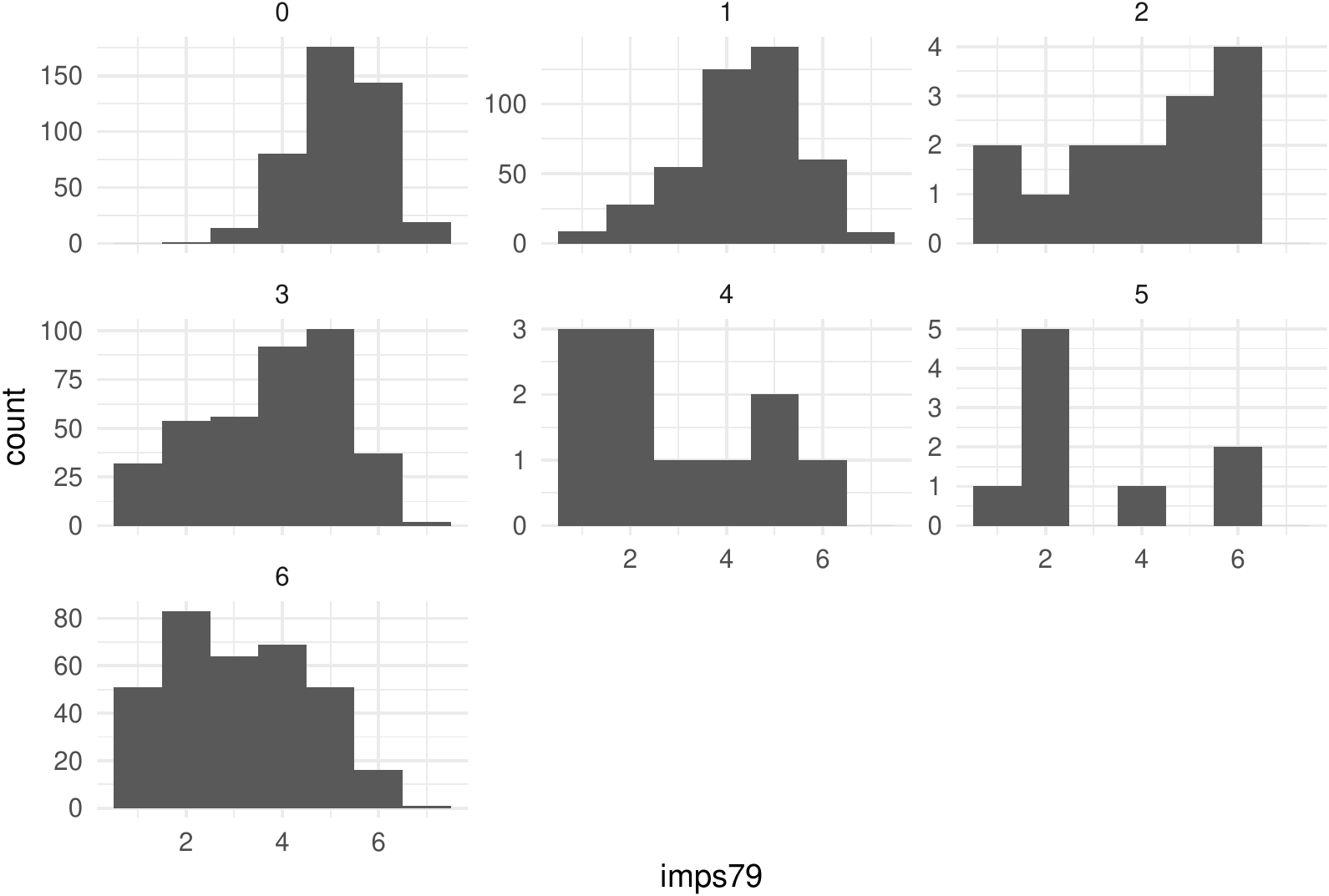} 

}

\caption{Distribution of the severity of illness ratings (IMPS item 79) at each week.}\label{fig:histograms-of-ratings-by-week}
\end{figure}

The analysis focuses on a single item of the Inpatient Multidimensional
Psychiatric Scale \citep{lorr1966inpatient}, which rates illness
severity on a scale from 1 (``normal'') to 7 (``among the most extremely
ill'').\footnote{The dataset provided contains some non-integer values
  for these ratings, presumably given to provide a finer-grained
  evaluation by the clinician.}. Most participants were measured on week
0 (99.31\%) and 1 (97.48\%), whilst the other main measurement points at
week 3 (85.58\%) and 6 (76.66\%) show more missing values. For a few
participants, ratings were instead obtained on week 2 (3.2\%), 4
(2.52\%), and/or 5 (2.06\%). Even when ignoring these rare deviations
from the main measurement points, there is a clear potential issue with
missing data and attrition, with 75.29\% being measured the intended
four times or more, and 15.1\% rated on just three occasions, and 9.61\%
only twice. The distribution of the ratings at each week is shown in
Figure \ref{fig:histograms-of-ratings-by-week}. As can be seen there,
ratings were generally relatively high at week 0, 1, and 3, but are
relatively lower at week 6. As there are only a small number of ratings
at week 2, 4, and 5, the empirical distributions for those weeks are
rather unreliable.

\begin{table}

\caption{\label{tab:glm-model-missingness}Results of a logistic regression analysis modelling missingness as a function of drug, week, and whether the week was a main measurement occasion or not.}
\centering
\begin{tabular}[t]{lrrrr}
\toprule
  & $\hat{\beta}$ & $\text{SE}(\hat{\beta})$ & $z$ & $P(>|z|)$\\
\midrule
\texttt{(Intercept)} & 1.921 & 0.393 & 4.884 & 0.000\\
\texttt{drug} & 0.433 & 0.463 & 0.936 & 0.349\\
\texttt{week} & 0.496 & 0.068 & 7.353 & 0.000\\
\texttt{main} & -5.381 & 0.382 & -14.103 & 0.000\\
$\texttt{drug} \times \texttt{week}$ & -0.112 & 0.081 & -1.395 & 0.163\\
$\texttt{drug} \times \texttt{main}$ & -0.596 & 0.446 & -1.335 & 0.182\\
\bottomrule
\end{tabular}
\end{table}

To gain initial insight into patterns underlying the missing data, we
modelled whether the IMPS rating was missing or not with a logistic
regression model. Predictors in the model were a dummy-coded variable
\texttt{drug} (placebo = 0, medicine = 1), \texttt{week} (from 0 to 6)
as a metric predictor, and a dummy-coded variable \texttt{main} to
indicate whether the rating was at a main measurement week (i.e.~at week
0, 1, 3, or 6). We also included an interaction between \texttt{drug}
and \texttt{week}, and between \texttt{drug} and \texttt{main}. The
results of this analysis (Table \ref{tab:glm-model-missingness}) show a
positive effect of \texttt{week} (such that missingness increases over
time), and a negative effect of \texttt{main}, with (many) more missing
values on weeks which are \emph{not} the main measurement weeks. The
positive effect of \texttt{week} is a clear sign of attrition. A
question now is whether this attrition is related to the severity of the
illness, in which case the ratings at week 6 would provide a biased view
on the true severity of illness after 6 weeks of treatment with a
placebo or medicine. There are of course different methods to assess
this, and many have been already applied to this particular dataset. Our
objective here is to incorporate a model of missingness into a hidden
Markov model, allowing missingness to depend on the latent state as well
as observable features such as the measurement week.

\hypertarget{hidden-markov-models-1}{%
\subsection{Hidden Markov models}\label{hidden-markov-models-1}}

We fitted HMMs in which we either assumed ratings are MAR, or assume
ratings are MNAR and state- and time-dependent. For each type of model
(MAR or MNAR), we fit versions with 2, 3, 4, or 5 states. Both types of
model assume \texttt{imps79}, the IMPS Item 79 ratings, follow a Normal
distribution, with a state-dependent mean and standard deviation. No
additional covariates were included, as the states are intended to
capture all the important determinants of illness severity. To model
effects of drug, we allow transitions between states, as well as the
initial state, to depend on \texttt{drug}. Whilst the initial
measurement at week 0 was made before administering the drug, we include
a possible dependence to account for any potential pre-existing
differences between the conditions. In the MNAR models, a second
(dichotomous) response variable \texttt{missing} is included, in
addition to \texttt{imps79}. The \texttt{missing} variable is modelled
with a logistic regression, using \texttt{week} and the dummy-coded
\texttt{main} variable as predictors, as these were found to be
important predictors in the (state-independent) logistic regression
analysis reported earlier. All models were estimated by maximum
likelihood using the EM algorithm implemented in depmixS4
\citep{depmixS4}.

\begin{table}

\caption{\label{tab:model-table}Modelling results for the MAR and MNAR hidden Markov models with 2-5 latent states.}
\centering
\begin{tabular}[t]{lrrrrr}
\toprule
model & \#states & log Likelihood & \#par & AIC & BIC\\
\midrule
MAR & 2 & -2422.675 & 16 & 4865.350 & 4919.146\\
 & 3 & -2266.603 & 30 & 4577.206 & 4695.558\\
 & 4 & -2225.871 & 48 & 4527.742 & 4732.168\\
 & 5 & -2182.390 & 70 & 4480.781 & 4792.799\\
\addlinespace
MNAR & 2 & -3074.628 & 22 & 6181.256 & 6267.330\\
 & 3 & -2889.040 & 39 & 5840.081 & 6006.849\\
 & 4 & -2841.108 & 60 & 5782.215 & 6051.197\\
 & 5 & -2800.336 & 85 & 5746.671 & 6139.385\\
\bottomrule
\end{tabular}
\end{table}

For both the MAR and MNAR models, the BIC indicates a three-state model
fits best, whilst the AIC indicates a five-state model (or higher) fits
best. Favouring simplicity, we follow the BIC scores here, and focus on
the three-state models.

We first consider the estimates of the MAR model. The estimated means
and standard deviations are \begin{equation*}
\gvc{\mu} = [2.315,4.339, 5.7] \quad \quad \gvc{\sigma} = [0.821,0.619, 0.567].
\end{equation*} Hence, the states are ordered, with state 1 being the
least severe, and state 3 the most severe. The prior probabilities of
the states, for treatment with placebo and drug respectively, are
\begin{equation*}
\gvc{\pi}_\text{placebo} = [0,0.333,0.667] \quad \quad \gvc{\pi}_\text{drug} = [0.005,0.307,0.689],
\end{equation*} and the transition probability matrices (with initial
states in rows and subsequent states in columns) are \begin{equation*}
\mathbf{T}_\text{placebo} = \left[ \begin{matrix} 0.963 &  0.004 &  0.032 \\ 0.118 &  0.878 &  0.004
\\ 0.027 &  0.046 &  0.927 \end{matrix} \right] \quad \quad \mathbf{T}_\text{drug} = \left[ \begin{matrix} 1 &  0 &  0 \\ 0.231 &  0.764 &  0.005
\\ 0.073 &  0.307 &  0.62 \end{matrix} \right].
\end{equation*} As expected, the initial state probabilities show little
difference between the treatments (as the initial measurement was
conducted before treatment commenced), but the transition probabilities
indicate that for those who were administered a real drug, transitions
to less severe states are generally more likely, indicating
effectiveness of the drugs. This is particularly marked for the most
severe state, where the probability of remaining in that state is 0.927
with a placebo, but 0.62 with a drug.

We next consider the three-state MNAR model. The means and standard
deviations are \begin{equation*}
\gvc{\mu} = [2.325,4.424, 5.756] \quad \quad \gvc{\sigma} = [0.833,0.668, 0.547]
\end{equation*} showing the same ordering of states in terms of
severity. The prior probabilities for placebo and drug groups are
\begin{equation*}
\gvc{\pi}_\text{placebo} = [0,0.393,0.607] \quad \quad \gvc{\pi}_\text{drug} = [0.004,0.349,0.647],
\end{equation*} and the transition probability matrices are
\begin{equation*}
\mathbf{T}_\text{placebo} = \left[ \begin{matrix} 0.93 &  0.005 &  0.065 \\ 0.123 &  0.872 &  0.005
\\ 0.026 &  0.031 &  0.942 \end{matrix} \right] \quad \quad \mathbf{T}_\text{drug} = \left[ \begin{matrix} 1 &  0 &  0 \\ 0.239 &  0.761 &  0.001
\\ 0.073 &  0.331 &  0.596 \end{matrix} \right].
\end{equation*} These estimates are close to those of the MAR model,
indicating little initial difference between the conditions, but
effectiveness of the drugs in the transition probabilities, which are
higher towards the less severe states than for the placebo condition.

\begin{table}

\caption{\label{tab:conditional-missingness-parameters}Parameter estimates of the state dependent logistic regression models for missingness, with lower and upper reflecting the lower and upper bounds of the approximate $95\%$ confidence intervals.}
\centering
\begin{tabular}[t]{llrrr}
\toprule
state & parameter & estimate & lower & upper\\
\midrule
1 & (Intercept) & 2.635 & 1.998 & 3.272\\
 & week & 0.149 & 0.028 & 0.269\\
 & main & -4.511 & -5.037 & -3.984\\
\addlinespace
2 & (Intercept) & 7.976 & 1.261 & 14.691\\
 & week & -0.510 & -1.978 & 0.959\\
 & main & -13.014 & -20.014 & -6.014\\
\addlinespace
3 & (Intercept) & 1.108 & 0.303 & 1.913\\
 & week & 0.710 & 0.538 & 0.883\\
 & main & -5.029 & -5.822 & -4.235\\
\bottomrule
\end{tabular}
\end{table}

\begin{figure}

{\centering \includegraphics[width=0.6\linewidth]{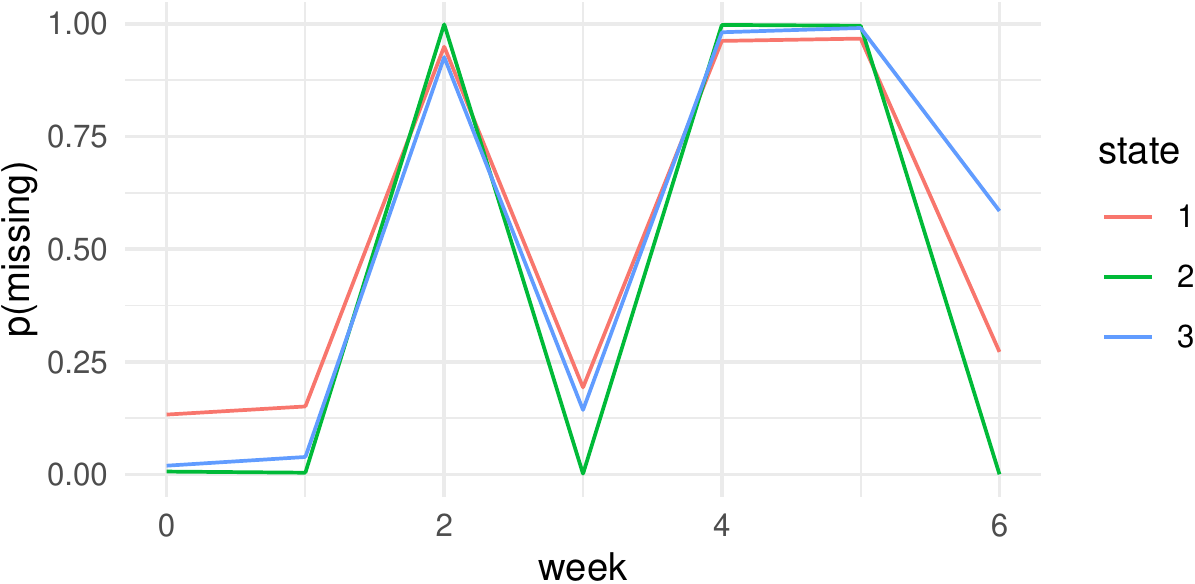} 

}

\caption{Predicted probability of missing IMPS Item 79 ratings by week for each state in the three-state MNAR hidden Markov model.}\label{fig:hmm-missing-prediction}
\end{figure}

Results of the state-dependent models for missingness are provided in
Table \ref{tab:conditional-missingness-parameters}. For all three
states, the confidence interval for the effect of \texttt{main} excludes
0, indicating a significantly lower proportion of missing ratings at the
main measurement weeks. In state 1 and 3, the confidence interval for
the effect of \texttt{week} also excludes 0, indicating a higher rate of
missing ratings over time, possibly due to attrition. For state 2, the
effect of \texttt{week} is not significant. Figure
\ref{fig:hmm-missing-prediction} depicts the predicted probability of
missing ratings for each state and week. This shows that in state 2, the
chance of missing data on the main measurement weeks is small at
\(p(M_t|S_t = 2)=0.003\), while it is high at \(p(M_t|S_t = 2) = 0.997\)
on the other weeks. In the other states, the probabilities are less
extreme, with missing (and non-missing) data occurring on the main
measurement weeks and the other weeks as well. In the final week 6,
those in the most severe state 3 are the most likely to have missing
data with \(p(M_t|S_t = 3) = 0.585\). For those in the least severe
state 1, the probability of missingness in week 6 is also substantial at
\(p(M_t|S_t = 1) = 0.272\).

Disregarding the modelling of missingness, the parameters of the MAR and
MNAR model seem reasonably close. This could be an indication that
missingness is independent of the hidden states and data are possibly
MAR. The likelihood of the MAR is not directly comparable to that of the
MNAR model, as the latter is defined over two variables (the rating and
the binary \texttt{missing} variable), while the former involves just a
single variable. However, we can test for equivalence by fitting a
constrained version of the MNAR model, where the parameters of the
missingness model are constrained to be identical over the states.
Unlike the MAR model, this restricted version of the MNAR model accounts
for patterns of missingness, allowing these to depend on \texttt{week}
and \texttt{main}, but not on the hidden state. A likelihood ratio test
indicates that this restricted model fits less well,
\(\chi^2(6) = 1328.57\) \(p < .001\). Hence, there is evidence that the
MNAR model is preferable to the MAR model and that missingness is indeed
state-dependent.

\begin{figure}

{\centering \includegraphics[width=0.7\linewidth]{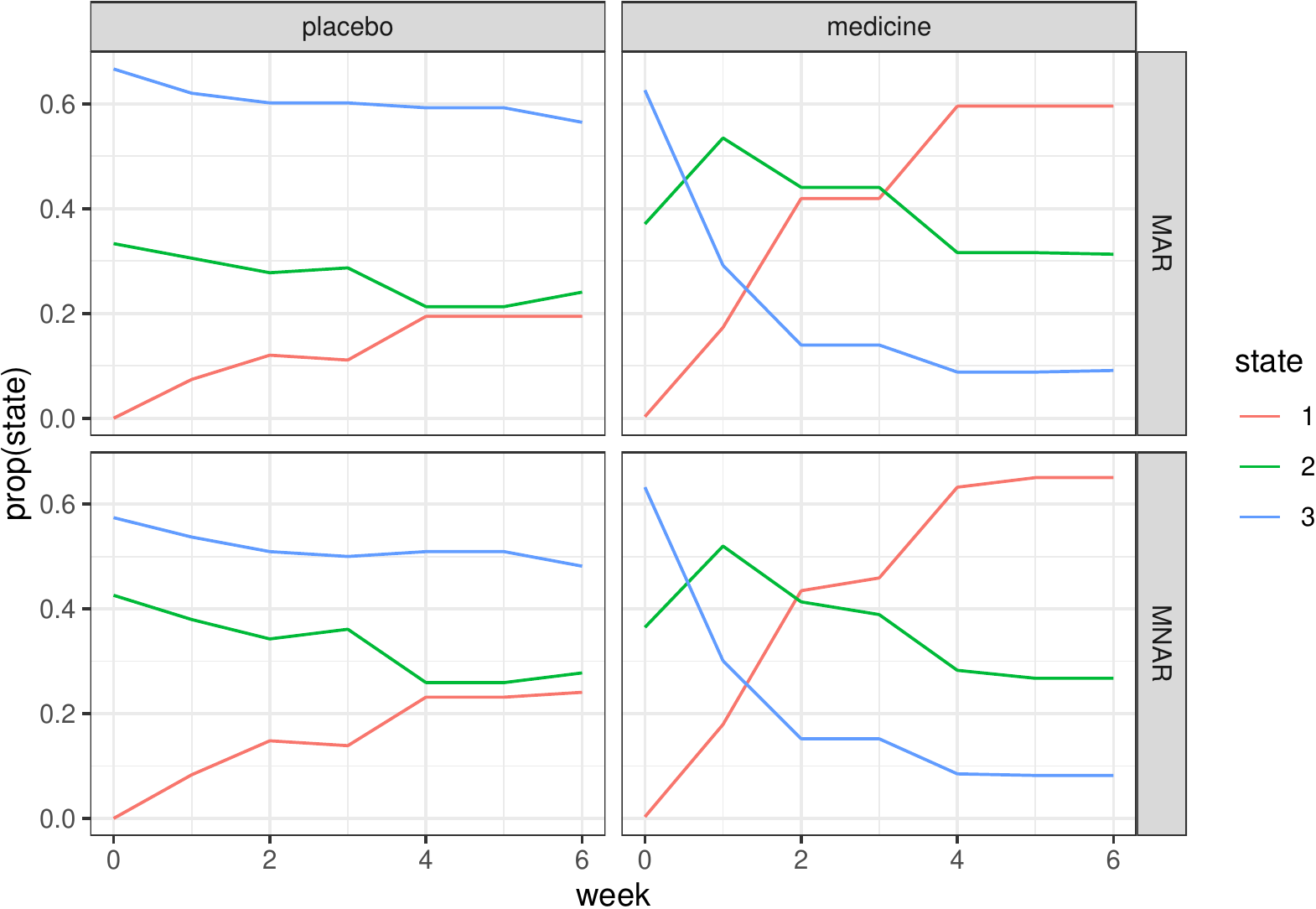} 

}

\caption{Proportions of maximum a posteriori (MAP) state assignments over weeks for the medication and placebo groups, according to the MAR and MNAR model.}\label{fig:imps74-posterior-state-plots}
\end{figure}

Whilst the MAR and MNAR model provide roughly equivalent parameters for
the severity ratings in the three states, when comparing the maximum a
posteriori (MAP) state classifications from the Viterbi algorithm
(Figure \ref{fig:imps74-posterior-state-plots}), we see that state
classifications for the the MAR model tend to be for the more severe
states. According to the MNAR model, during the main measuring weeks,
missing values are relatively likely in the least severe state 1. Hence,
those with missing values are more likely to be assigned to the least
severe state. This is in line with the analysis of
\citet{hedeker1997application}, who found evidence that dropouts in the
medication condition showed more improvement in their symptoms.

It is worthwhile to note that the MAP states are also determined for
time points with missing data, as the transition probabilities make
certain states more probable than others, even when there is no direct
measurement available. This provides a potentially meaningful basis to
impute missing values, with another option being expected rating as a
function of the posterior probability of all states. As imputation is
not the focus of this study, we leave this to be investigated in future
work.

\hypertarget{discussion}{%
\section{Discussion}\label{discussion}}

Previous work on missing data in hidden Markov models has mostly
focussed on cases where missing values are assumed to be missing at
random (MAR). Here, we addressed situations where data is missing not at
random (MNAR), and missingness depends on the hidden states. Simulations
showed that including a submodel for state-dependent missingness in a
HMM is beneficial when missingness is indeed state-dependent, whilst
relatively harmless when data is MAR. However, when the form of
state-dependent missingness is misspecified (e.g.~the effect of
measurable covariates on missingness ignored), results may be biased. In
practice, it is therefore advisable to consider the potential effect of
covariates in the state-dependent missingness models. A reasonable
strategy is to first model patterns of missingness through e.g.~logistic
regression, and then include important predictors from this analysis
into the state-dependent missingness models. Applying this strategy to a
real example about severity of schizophrenia in a clinical trial with
substantial missing data, we showed that assuming data is MAR may lead
to possible misclassification of patients to states (towards more severe
states in this example). Whilst the ground truth is unavailable in such
real applications, model comparison can be used to justify a
state-dependent missingness model. Using flexible analysis tools such as
the depmixS4 package \citep{depmixS4} makes specifying, estimating, and
comparing hidden Markov models with missing data specifications
straightforward. There is then little reason to ignore potentially
non-ignorable patterns of missing data in hidden Markov modelling.

Another approach to dealing with non-ignorable missingness (MNAR) is the
pattern-mixture approach of Little
\citetext{\citeyear{little1993pattern}; \citeyear{little1994class}}. The
main idea of this approach is to group units of observations
(e.g.~patients) by the pattern of missing data, and allowing the
parameters of a statistical model for the observations \(Y\) to
dependent on the missingness \emph{pattern} \(M_{1:\nT}\). There are
certain similarities between this approach and modelling missingness as
state-dependent. Rather than conditionalizing on a pattern of missing
values, a hidden Markov model conditionalizes on a pattern (sequence) of
hidden states, \(\fstate_{1:\nT}\), and the marginal distribution of the
observations is effectively a multivariate mixture \begin{equation}
\dens(\obs_{1:\nT} | \modpar) = \sum_{\fstate_{1:\nT} \in \mathcal{S}^\nT}  \sum_{m_{1:\nT} \in \mathcal{M}^\nT} \dens(\obs_{1:\nT} | m_{1:\nT} , \fstate_{1:\nT}, \modpar) \dens(m_{1:\nT} | \fstate_{1:\nT}, \modpar) \dens(\fstate_{1:\nT} | \modpar)
\end{equation} (note that \(\modpar\) here includes all parameters, so
\(\phi\)). A pattern-mixture model would instead propose
\begin{equation}
\dens(\obs_{1:\nT} | \modpar) = \sum_{m_{1:\nT} \in \mathcal{M}^\nT} \dens(\obs_{1:\nT} | m_{1:\nT}, \modpar) \dens(m_{1:\nT} | \modpar) .
\end{equation} Trivially, if we set the number of hidden states to
\(\ns = 1\), both models are the same. Another trivial equivalence is
through a one-to-one mapping between \(m_{1:\nT}\) and
\(\fstate_{1:\nT}\). This could be obtained of by setting \(\ns = 2\),
assuming the Markov process is of order \(\nT\), and fixing
e.g.~\(\Prob(M_{t} = 0|\state = 1) = 1\) and
\(\Prob(M_{t} = 1|\state = 2) = 1\). More interesting is to investigate
cases where the procedures are similar, but not necessarily equivalent.
The general pattern-mixture model is often underidentified
\citep{little1993pattern}. For time-series of length \(\nT\), there are
\(2^\nT\) possible missing data patterns. Without further restrictions,
estimating the mean vectors and covariance matrices for all these
components is not possible, due to the structural missing data in those
patterns. The state-dependent MNAR hidden Markov model is identifiable
insofar as the HMM for the observed variable \(\obs\) is identifiable.
It is convenient, but not necessary, to assume a first-order Markov
process. Higher-order Markov processes may allow the model to capture
complex effects of missingness. Another option is to use the missingness
indicator \(M_t\) as a covariate on initial and transition
probabilities, rather than a dependent variable. We leave investigation
of such alternative models to future work.

\bibliographystyle{agsm}
\bibliography{refs.bib}

\end{document}